\documentclass[3p,times]{elsarticle}

\usepackage{ecrc}
\usepackage{subfig}
\usepackage{wrapfig}

\volume{00}

\firstpage{1}

\journalname{Procedia Computer Science}

\runauth{E. Gaburov et al.}

\jid{procs}

\jnltitlelogo{Procedia Computer Science}

\usepackage{amssymb}

\usepackage{graphicx}
\usepackage{amssymb}

\usepackage[figuresright]{rotating}

\usepackage[]{fontenc}
\usepackage[latin1]{inputenc}
\usepackage{listings}

\newcommand{\fig}[1]{\mbox{Fig. #1}}

\begin{document}

\begin{frontmatter}



\dochead{International Conference on Computational Science, ICCS 2010}

\title{Gravitational tree-code on graphics processing units:
  implementation in CUDA}



\author[LU]{Evghenii Gaburov}
\ead{egaburov@strw.leidenuniv.nl}
\author[LU]{Jeroen B\'edorf}
\ead{bedorf@strw.leidenuniv.nl}
\author[LU]{Simon Portegies Zwart}
\ead{spz@strw.leidenuniv.nl}

\address[LU]{Leiden Observatory, Leiden University, Leiden The Netherlands }


\begin{abstract}
  We present a new very fast tree-code which runs on massively
  parallel Graphical Processing Units (GPU) with NVIDIA CUDA
  architecture. The tree-construction and calculation of multipole
  moments is carried out on the host CPU, while the force calculation
  which consists of tree walks and evaluation of interaction list is
  carried out on the GPU. In this way we achieve a sustained
  performance of about 100GFLOP/s and data transfer rates of about
  50GB/s. It takes about a second to compute forces on a million
  particles with an opening angle of $\theta \approx 0.5$. The code has
  a convenient user interface and is freely available for
  use\footnote{{\tt http://castle.strw.leidenuniv.nl/software/octgrav.html}}.
\end{abstract}

\begin{keyword}
\sep gravity
\sep tree-code
\sep GPU
\sep parallel
\sep CUDA


\end{keyword}

\end{frontmatter}





\section{Introduction}\label{sect:introduction}

Direct force evaluation methods have always been popular because of
their simplicity and unprecedented accuracy.  Since the mid 1980's,
however, approximation methods like the hierarchical tree-code
\citep{1986Natur.324..446B} have gained enormous popularity among
researchers, in particular for studying astronomical self-gravitating
$N$-body systems \citep{2003gnbs.book.....A} and for studying
soft-matter molecular-dynamics problems \citep{Frenkel2001}. For
these applications, direct force evaluation algorithms strongly limit
the applicability, mainly due to the $O(N^2)$ time complexity of the
problem.

Tree-codes, however, have always had a dramatic set back compared to
direct methods, in the sense that the latter benefits from the
developments in special purpose hardware, like the GRAPE and MD-GRAPE
family of computers \cite{1998sssp.book.....M,2001ASPC..228...87M},
which increase workstation performance by two to three orders of
magnitude. On the other hand, tree-codes show a better scaling of the
compute time with the number of processors on large parallel
supercomputers
\cite{169640, WarBecGod97} compared to direct $N$-body methods
\cite{2007NewA...12..357H, 1233038}. As a results, large
scale tree-code simulations are generally performed on Beowulf-type
clusters or supercomputers, whereas direct $N$-body simulations are
performed on workstations with attached GRAPE hardware.

Tree-codes, due to their hierarchical and recursive nature are hard to
run efficiently on dedicated Single Instruction Multiple Data (SIMD)
hardware like GRAPE, though some benefit 
has been demonstrated by using
pseudo-particle methods to solve for the higher-order moments in the
calculation of multipole moments of the particle distributions in
grid cells \cite{2004ApJS..151...13K}.

Recently, the popularity of computer games has led to 
the development of massively parallel vector processors for rendering 
three-dimensional graphical images. Graphical Processing
Units (or GPUs) have evolved from fixed function hardware to general
purpose parallel processors. The theoretical peak speed of these
processors increases at a rate faster than Moores' law
\citep{Moore}, and at the moment top roughly 200\,GFLOP for a single
card. The cost of these cards is dramatically reduced by the enormous
volumes in which they are produced, mainly for gamers, whereas
GRAPE hardware remains relatively expensive.


The gravitational $N$-body problem proved to be rather ideal to port
to modern GPUs, and the first successes in porting the $N$-body
problem to programmable GPUs were achieved by \cite{Nyland04}, but it
was only after the introduction of the NVIDIA G80 architecture
that accurate force evaluation algorithms could be implemented
\cite{2007NewA...12..641P} and that the performance became comparable
to special purpose computers \cite{2008NewA...13..103B, Gaburov2009630}.

Even in these implementations, the tree-code, though pioneered in
\cite{2008NewA...13..103B}, still hardly resulted in a speed-up
compared to general purpose processors.  In this paper we present a
novel implementation of a tree-code on the NVIDIA GPU hardware using
the CUDA programming environment.

\section{Implementation}\label{sect:implementation}
\vspace{-7pt}
\begin{wrapfigure}{l}{0.35\textwidth}
  \centering
  \includegraphics[scale=0.32]{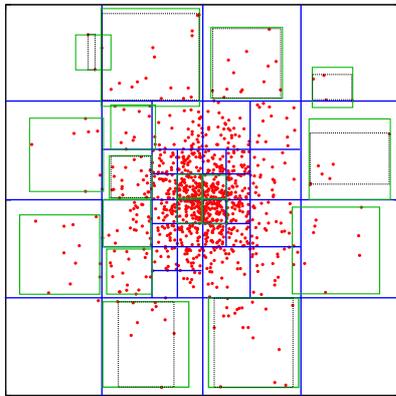}
  \vspace{-10pt}
  \caption{\small Illustration of our tree-structure, shown in 2D
    for clarity. Initially, the space is recursively subdivided into 
    cubic cells until all cells contain less than $N_{\rm leaf}$
    particles (blue squares). All cells (including parent cells) are
    stored in a tree-structure. Afterwards, we compute a tight
    bounding box for the particles in each cell (dotted rectangles)
    and cell's boundary. The latter is a cube with a side length equal
    to the largest side length of the bounding box and the same centre
    (green squares). }
  \label{fig:octree}
  \vspace{-10pt}
\end{wrapfigure}
In the classical implementation of the tree-code algorithm all the work 
is done on the CPU, since special purpose hardware was not available at
that time \cite{1986Natur.324..446B}.  With the introduction of GRAPE
special purpose hardware \cite{1990CoPhC..60..187I,
  1991PASJ...43..841F}, it became computationally favourable to let
the special purpose hardware, instead of the CPU, calculate accelerations.
Construction of the interaction list in these implementations takes
nearly as much time as calculating the accelerations. Since the latest
generation of GPUs allows more complex operations, it becomes possible
to build the interaction list directly on the GPU. In this case, it is
only necessary to transport the tree-structure to the GPU. Since the
bandwidth on the host computer is about an order of magnitude lower
than on the GPU, it is also desirable to offload bandwidth intensive
operations to the GPU.  The construction of the interaction list is
such an operation. The novel element in our tree-code is construction
of the interaction list on the GPU. The remaining parts of the
tree-code algorithm (tree-construction, calculation of node properties
and time integration) are executed on the host. The host is also
responsible for the allocation of memory and the data transfer to and
from the GPU. In the next sections we will cover the details of the
host and device steps.

\subsection{Building the octree}

We construct the octree in the same way as done in the original BH
tree-code. We define the computational domain as a cube containing all
particles in the system. This cube is recursively divided into eight
equal-size cubes called cells. The length of the resulting cells is
half the length of the parent cell. Each of these cells is
further subdivided, until less than $N_{\rm leaf}$ particles are
left. We call these cells leaves, whereas cells containing more than 
$N_{\rm  leaf}$ particles are referred to as nodes. The cell containing 
all particles is the root node. 

The resulting tree-structure is
schematically shown in \fig{\ref{fig:octree}}. From this
tree-structure we construct groups for the tree walk (c.f. 
section \ref{sect:listconst}), which are the cells with the number of 
particles less than
$N_{\rm groups}$, and compute properties for each cell, such as
boundary, mass, centre of mass, and quadrupole moments, which are
required to calculate accelerations \cite{1993ApJ...414..200M}.

\begin{wrapfigure}{r}{0.5\textwidth}
  \centering
  \includegraphics[scale=0.57]{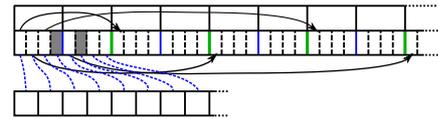}
  \vspace{-10pt}
  \caption{\small Illustration of the tree structure as stored in
    device memory.}
  \label{fig:tree_memory}
  \vspace{-10pt}
\end{wrapfigure}

In order to efficiently walk the octree on the device, its structure
requires some reorganisation.
In particular, we would like to minimise the number of memory accesses
since they are relative expensive (up to 600 clock cycles).
In \fig{\ref{fig:tree_memory}}, we show the tree-structure
as stored on the GPU. The upper array in the figure is the link-list
of the tree, which we call the main tree-array. Each element in this
array (separated by blue vertical lines) stores four integers in a
single 128-bit words (dashed vertical lines). This structure is
particularly favourable because the device is able to read a 128-bit
word into four 32-bit registers using one memory access instruction. 
Two array-elements represent one cell in the tree (green line) with
indices to each of the eight children in the main tree-array
(indicated by the arrows). A grey filled element in this list means
that a child is a leaf (it has no children of its own), and hence it
needs not to be referenced. We also use auxiliary tree-arrays in the
device memory which store properties of each cell, such as its
boundary, mass, centre of mass and multiple moments.  The index of
each cell in the main tree-array is directly related to its index
in the auxiliary tree-arrays by bitshift and addition operations.

The device execution model is designed in such a way that threads
which execute the same operation are grouped in warps, where each warp
consists of 32 threads. Therefore, all threads in a warp follow the
same code path. If this condition is not fulfilled, the divergent code
path is serialised, therefore negatively impacting the performance
\citep{NVIDIA.CUDA}. To minimise this, we rearrange groups in 
memory to make sure that neighbouring groups in space are also
neighbouring in memory. Combined with similar tree paths that
neighbouring groups have, this will minimise data and code path
divergence for neighbouring threads, and therefore further improves
the performance.

\subsection{Construction of an interaction list}\label{sect:listconst}
In the standard BH-tree algorithm, the interaction lists are
constructed for each particle, but particles in the same groups have
similar interaction lists. We make use of this fact by building the
lists for groups instead of particles \citep{1990JCoPh..87..161B}. The
particles in each group, therefore, share the same interaction list,
which is typically longer than it would have been by determining it on
a particle basis. The advantage here is the reduction of the number of
tree walks by $N_{\rm group}$. The tree walk is performed on the GPU
in such a way that a single GPU thread is used per group. To take 
advantage of the cached texture memory, we make sure that neighbouring
threads correspond to neighbouring groups.

Owing to the massively parallel architecture of the GPU, two tree
walks are required to construct interaction lists. In the first walk,
each thread computes the size of the interaction list for a
group. This data is copied to the host, where we compute the total
size of the interaction list, and memory addresses to which threads
should write lists without intruding on other threads' data. In the
second tree walk, the threads write the interaction lists to the
device memory.

\begin{lstlisting}[ caption = {\small A pseudo code for our non-recursive
    stack-based tree walk.}, label=list:tree_walk ]
 while (stack.non_empty) 
   node = stack.pop                      ;; get next node from the stack
   one  = fetch(children, node + 0)      ;; cached fetch 1st four children 
   two  = fetch(children, node + 1)      ;; cached fetch 2nd four children 
   test_cell<0...4>(node, one, stack)    ;; test sub-cell in octant one to four
   test_cell<5...8>(node, two, stack)    ;; test sub-cell in octant four to eight
\end{lstlisting}

\begin{lstlisting}[ caption = {\small Pseudo code for {\tt test\_cell}
    subroutine.}, label=list:test_cell]
template<oct>test_cell(node, child, stack) 
   child = fetch(cell_pos, 8*node +oct)    ;; fetch data of the child   
   if (open_node(leaf_data, child))        ;; if the child has to be opened, 
     if (child != leaf) stack.push(child)  ;;  store it in stack if it is a node
     else              leaf += 1           ;;  otherwise increment the leaf counter 
   else                cell += 1           ;; else, increment the cell counter
\end{lstlisting}

We implemented the tree walk via a non-recursive stack-based algorithm
(the pseudo code is shown in List \ref{list:tree_walk}), because the
current GPU architecture does not support recursive function calls. In
short, every node of the tree, starting with the root node, reads
indices of its children by fetching two consecutive 128-bit words
(eight 32 bit integers) from texture memory.  Each of these eight
children is tested against the node-opening criteria $\theta$
(the pseudo code for
this step is shown in List \ref{list:test_cell}), and in the case of a
positive result a child is stored in the stack (line 4 in the listing),
otherwise it is considered to be a part of the interaction list. In
the latter case, we check whether the child is a leaf, and if so, we
increment a counter for the leaf-interaction list (line 5), otherwise
a counter for the node-interaction list (line 6). This division of the
interaction lists is motivated by the different methods used to
compute the accelerations from nodes and leaves 
(c.f. section \ref{sect:calcacc}).
In the second tree walk, we store the index of the cell in the 
appropriate interaction list instead of counting the nodes and leafs.

\subsection{Calculating accelerations from the interaction list}
\label{sect:calcacc}
\begin{figure*} 
  \centering
  \includegraphics[width=\columnwidth]{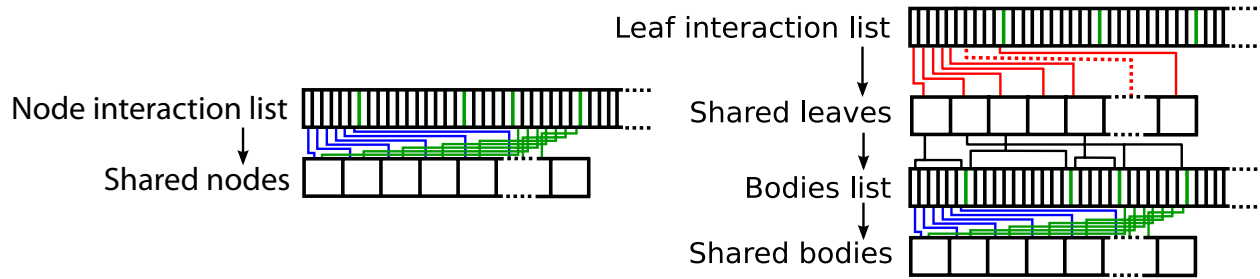}
  \vspace{-15pt}
  \caption{\small Memory access pattern in a body-node (left) and body-leaf
    (right) interaction.}
  \vspace{-15pt}
  \label{fig:body_node_leaf}
\end{figure*}

In the previous step, we have obtained two interaction lists: one for
nodes and one for leaves. The former is used to compute accelerations
due to nodes, and the latter due to leaves.  The pseudo-code for a
particle-node interaction is shown in List \ref{list:body-cell} and
the memory access pattern is demonstrated in the left panel of
\fig{\ref{fig:body_node_leaf}}. This algorithm is similar to the one
used in the {\tt kirin} and {\tt sapporo} libraries for direct
$N$-body simulations \cite{2008NewA...13..103B, Gaburov2009630}. In
short, we use a block of threads per group, such that a thread in a
block is assigned to a particle in a group; these particles share the
same interaction list. Each thread loads a fraction of the nodes from the
node-interaction list into shared memory (blue threads in the figure,
lines 2 and 3 in the listing). To ensure that all the data is loaded
into shared memory, we put a barrier for all threads (line 4), and
afterwards each thread computes gravitational acceleration from the
nodes in shared memory (line 5). Prior loading a new set of nodes into
the shared memory (green threads in the figure), we ensure that all
the threads have completed their calculations (line 6). We repeat this
cycle until the interaction list is exhausted.

\begin{lstlisting}[caption = {\small Body-node interaction},
  label=list:body-cell]
 for (i = 0; i < list_len; i += block_size)   
   cellIdx = cell_interact_lst[i + thread_id]
   shared_cells[threadIdx] = cells_lst[cellIdx] ;; read nodes to the shared memory
   __syncthreads()                              ;; thread barrier
   interact(body_in_a_thread, shared_cell)      ;; evaluate accelerations
   __syncthreads()                              ;; thread barrier
\end{lstlisting}

\begin{lstlisting}[caption = {\small Body-leaf interaction}, label=list:body-leaf]
 for (i = 0; i < list_len; i += block_size) {
   leaf = leaf_interaction_list[i + threads_id]  
   shared_leaves[threadIdx] = cells_list[leaf]  ;; read leaves to the shared memory
   __syncthreads()                             
   for (j = 0; j < block_size; j++)            ;; process each leaf 
     shared_bodies[thread_id] =  bodies[shared_leaves[j].first + thread_id]   
     __syncthreads();    
     interact(body_in_a_thread, shared_bodies, shared_leaves[j].len);
     __syncthreads();
\end{lstlisting}

Calculations of gravitational acceleration due to the leaves differs
in several ways. The pseudo-code of this algorithm is presented in
List \ref{list:body-leaf}, and the memory access pattern is displayed
in the right panel of Fig.  \ref{fig:body_node_leaf}. First, each
thread fetches leaf properties, such as index of the first body and
the number of bodies in the leaf, from texture memory into 
shared memory (red lines in the figure, lines 2 and 3 in the
listing). This data is used to identify bodies from which the
accelerations have to be computed (black lines). Finally, threads read
these bodies into shared memory (blue and green lines, line 6) in
order to calculate accelerations (line 8). This process is repeated
until the leaf-interaction list is exhausted.

\section{Results}\label{sect:results}
\vspace{-7pt}
In this section we study the accuracy and performance of the tree
code.  First we quantify the errors in acceleration produced by the
code and then we test its performance. For this purpose we use a model
of the Milky Way galaxy \citep{2005ApJ...631..838W}. We model the
galaxy with $N=10^4, 3\cdot 10^4, 10^5, 3\cdot10^5, 10^6, 3\cdot10^6$
and $10^7$ particles, such that the mass ratio of bulge, disk and halo
particles is 1:1:4. We then proceed with the measurements of the code
performance. In all test we use $N_{\rm leaf} = 64$ and $N_{\rm
  group} = 64$ which we find produce the best performance on both
G80/G92 and GT200 architecture. The GPU used in all the tests is a
GeForce 8800Ultra.

\begin{figure*}
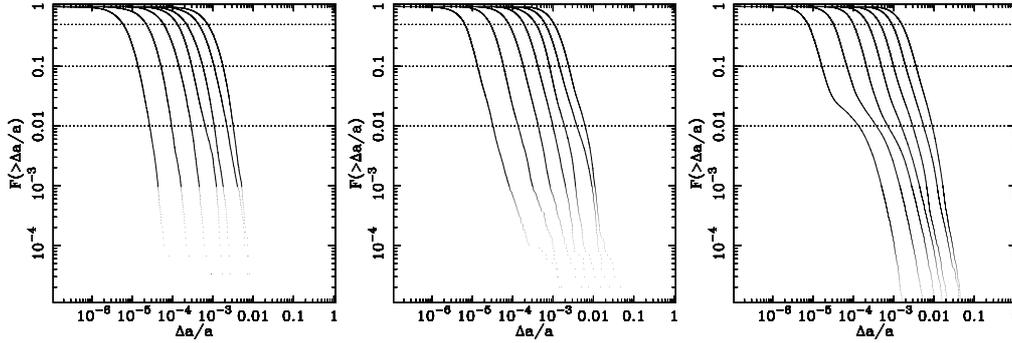

  \centering
  \includegraphics[scale=0.25]{plots/err_mw_0030K}$\,$
  \includegraphics[scale=0.25]{plots/err_mw_0100K}$\,$
  \includegraphics[scale=0.25]{plots/err_mw_1000K}
  \vspace{-10pt}
  \caption{\small Each panel displays a fraction of particles having relative
    acceleration error (vertical axis) greater than a given value
    (horizontal axis). In each panel, we show errors for various
    opening angles from $\theta = 0.2$ (the leftmost curve in each
    panel), $0.3$, $0.4$, $0.5$, $0.6$ and $0.7$ (the rightmost
    curve). The number of particles are $3\cdot 10^4$, $10^5$, $10^6$
    for panels from left to right respectively. The dotted horizontal
    lines show 50\%, 10\% and 1\% of the error distribution.}
  \label{fig:err_dist}
  \vspace{-10pt}
\end{figure*}

\subsection{Accuracy of approximation}
We quantify the error in acceleration in the following way: $ \Delta
a/a = |{\bf a_{\rm tree} - a_{\rm direct}}|/|{\bf a_{\rm direct}}|$,
where ${\bf a}_{\rm tree}$ and ${\bf a}_{\rm direct}$ are
accelerations calculated by the tree and direct summation
respectively. The latter was carried out on the same GPU as the
tree-code. This allowed us to asses errors on systems as large as 10
million particles\footnote{We used the NVIDIA 8800Ultra GPU for this
  paper, and it takes $\sim$10 GPU hours to compute the exact force on a
  system with 10 million particles with double precision emulation
  \citep{Gaburov2009630}}. In \fig{\ref{fig:err_dist}} we show error
distributions for different numbers of particles and for different
opening angles. In each panel, we show which fraction of particles
(vertical-axis) has a relative error in acceleration larger than a
given value (horizontal axis). The horizontal lines show 50th, 10th
and 1st percentile of cumulative distribution. This data shows that
acceleration errors in this tree-code are consistent with the errors
produced by existing tree-codes with quadrupole corrections
\citep{2002JCoPh.179...27D, 2001NewA....6...79S, 2001PhDT........21S}.

\begin{figure*}
  \centering
  \includegraphics[width=\columnwidth]{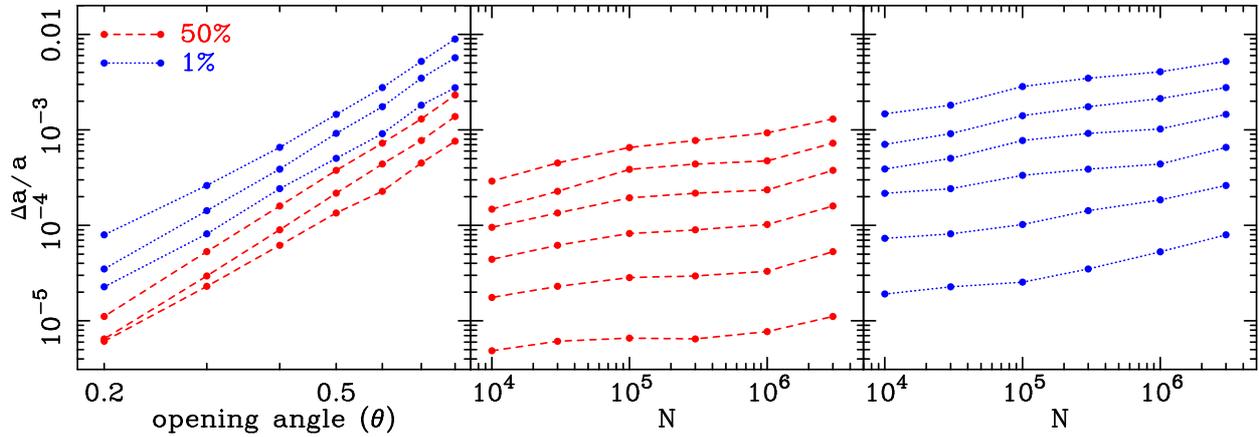}
  \vspace{-20pt}
  \caption{\small The median and the first percentile of the relative
    acceleration error distribution as a function of the opening angle
    and the number of particles. In the leftmost panel we show lines for
    $3\cdot 10^4$ (the bottom dotted and dashed lines) and $3\cdot
    10^6$ (the top dotted and dashed lines) particles. The middle and
    the right panels display the error for $\theta= 0.2$ (the bottom
    lines), $0.3, 0.4, 0.5, 0.6$ and $0.7$ (the upper lines).}
  \label{fig:errors_data}
  \vspace{-10pt}
\end{figure*}

We show dependence of errors on both opening angle and number of
particles in \fig{\ref{fig:errors_data}}. In the leftmost panel of the
figure, we plot the median and the first percentile of the relative
force error distribution as a function of the opening angle $\theta$
for various number of particles $N=3\cdot 10^4$ (the lowest blue
dotted and red dashed lines), $3\cdot 10^5$ and $3\cdot 10^6$ (the
upper blue dotted and red dashed lines). As expected, the error
increases as a function of $\theta$ with the following scaling from the
least-squared fit, $\Delta a/a \propto \theta^4$. However, the errors
increase with the number of particles: the error doubles when the
number of particles is increased by two orders of magnitude. This
increase of the error is caused by the large number of particles in a
leaf, which in our case is 64, to obtain the best performance. We
conducted a test with $N_{\rm leaf} = 8$, and indeed observed the
expected decrease of the error when the number of particles increases;
this error, however, is twice as large compared to $N_{\rm leaf} = 64$
for $N \sim 10^6$.

\subsection{Timing}

\begin{figure*}
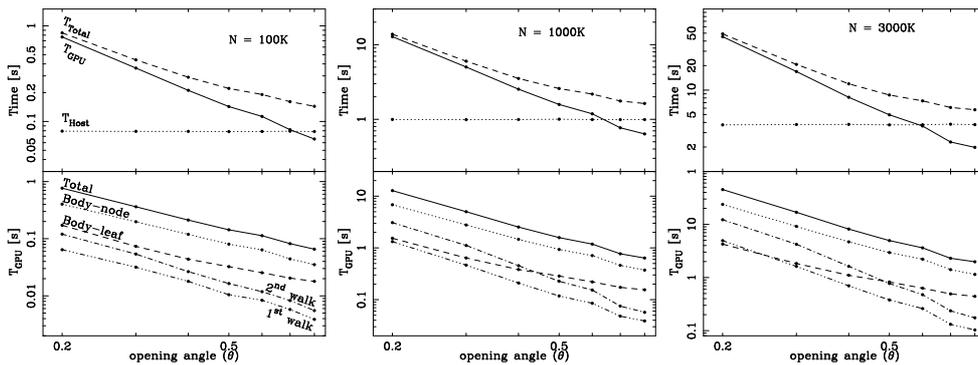

  \centering
  \includegraphics[scale=0.25]{plots/time_mw_0100K}$\,$
  \includegraphics[scale=0.25]{plots/time_mw_1000K}$\,$
  \includegraphics[scale=0.25]{plots/time_mw_3000K}
  \vspace{-10pt}
  \caption{\small Wall-clock timing results as function of the opening angle
    and number of particles. In each panel, the solid line shows the
    time spent on the GPU. The dotted line on the top panel shows the
    time spent on the host, and the total wall-clock time is shown
    with the dashed line.}
  \label{fig:timing}
  \vspace{-5pt}
\end{figure*}
\begin{figure}
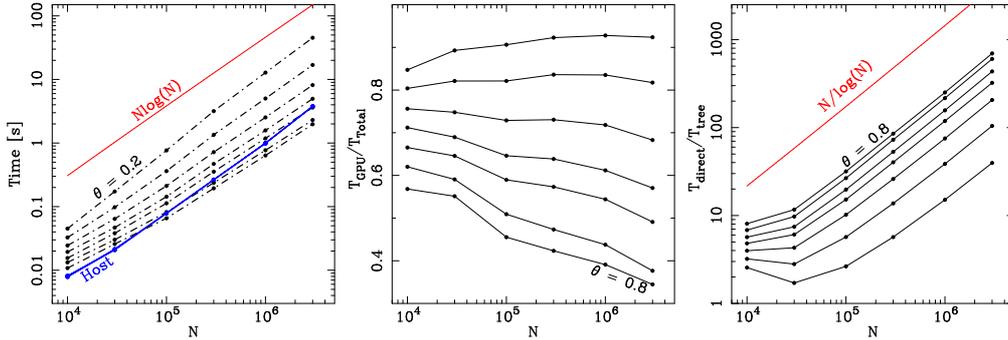

  \centering
  \includegraphics[scale=0.25]{plots/timing}$\,$
  \includegraphics[scale=0.25]{plots/time_ratio}$\,$
  \includegraphics[scale=0.25]{plots/direct_tree}
  \vspace{-10pt}
  \caption{\small Timing results as a function of particle number. The
    leftmost panel displays time spent on the GPU (black dash-dotted
    lines) and host CPU (blue solid line) parts as a function of the
    number of particles. The expected scaling $N\log(N)$ is shown
    in the red solid line. The ratio of the time spent on GPU to
    the total wall-clock time is given in the middle panel. The
    speed-up compared to direct summation is shown in the rightmost
    panel. The expected scaling $N/\log(N)$ is shown with a solid red line.}
  \label{fig:time_ratio}
  \vspace{-10pt}
\end{figure}

In \fig{\ref{fig:timing}} we present the timing data as a function of
$\theta$ and for various $N$. The $T_{\rm Host}$ (dotted line in the
figure) is independent of $\theta$, which demonstrates that
construction of the octree only depends on the number of particles in the
system, with $T_{\rm Host} \propto N\log N$. This time becomes
comparable to the time spend on the GPU calculating accelerations for
$N \gtrsim 10^6$ and $\theta \gtrsim 0.5$. This is caused by the
empirically measured near-linear scaling of time spend on GPU with
$N$. As the number of particles increases, the GPU part of the code
performs more efficiently, and therefore the scaling drops from $N
\log N$ to near-linear (\fig{\ref{fig:time_ratio}}). We therefore
conclude, that the optimal opening angle for our code is $\theta
\approx 0.5$.




In the leftmost panel of \fig{\ref{fig:time_ratio}} we show $N$
dependence of the time spent on the host and the device for various
opening angles. In particular, $T_{\rm GPU}$ scaling falls between
$N\log(N)$ and $N$, which we explained by the increased efficiency of
the GPU part of our code with larger number of particles. This plot
also shows that the host calculation time is a small fraction of the
GPU calculation time, except for $N \gtrsim 10^6$ and $\theta \gtrsim
0.5$. The middle panel of the figure shows the ratio of the time spent
on the device to the total time.
Finally, the rightmost 
panel shows the ratio between the time required to compute forces by 
direct summation and the time required by the tree-code. As we 
expected, the scaling is consistent with $N^2/(N\log(N)) = N/log(N)$. 


\subsection{Device utilisation}
\begin{figure*}
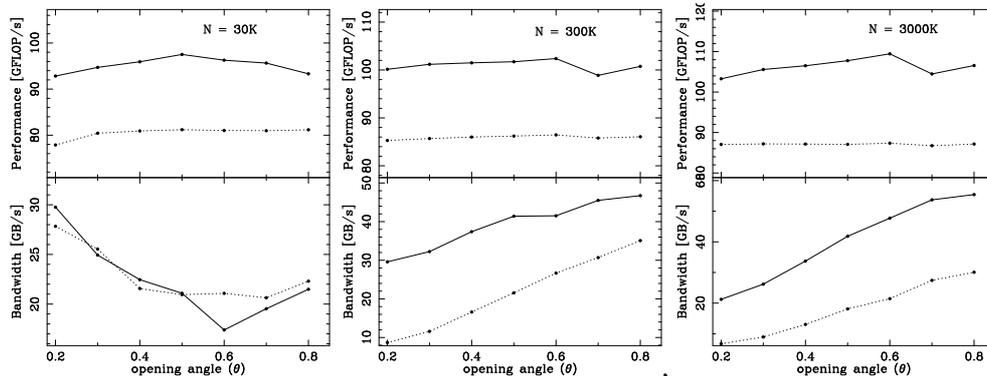

  \centering
  \includegraphics[scale=0.25]{plots/performance_mw_0030K}$\,$
  \includegraphics[scale=0.25]{plots/performance_mw_0300K}$,$
  \includegraphics[scale=0.25]{plots/performance_mw_3000K}
  \vspace{-10pt}
  \caption{\small Device utilisation as a function of the opening angle
    and number of particles. Each bottom panel shows the bandwidth 
    for the first tree walk (solid line) and the second tree walk
    (dotted line). The top halves show the performance of the
    calculation of the accelerations for the node interaction list
    (solid line) and the leaf interaction list (dotted line) in GFLOP/s.}
  \label{fig:performance}
  \vspace{-10pt}
\end{figure*}

We quantify the efficiency of our code to utilise the GPU resources by measuring both
instruction and data throughput, and then compare the results to the theoretical peak
values provided by the device manufacturer. In \fig{\ref{fig:performance}} we show both
bandwidth and computational performance as function of $\theta$ for three different
$N$. We see that the calculation of accelerations operates at about 100GFLOPs\footnote{We
  count 38 and 70 FLOPs for each body-leaf and body-node interaction respectively.}. This
is comparable to the peak performance of the GRAPE-6a special-purpose hardware, but this
utilises only $\sim 30$\% of the computational
power of the GPU\footnote{Our tests were carried out on a NVIDIA
  8800Ultra GPU, which has 128 streaming processors each operating at
  clock speed of 1.5Ghz. Given that the GPU is able to perform up to two
  floating point operation per clock cycle, the theoretical peak performance
  is $2\times 128 = 384$GFLOP/s.}. This occurs because the average
number of bodies in a group is a factor of 3 or 4 smaller than the
$N_{\rm group}$, which we set to 64 in our tests. On average,
therefore, only about 30\% of the threads in a block are
active.

The novelty of this code is the GPU-based tree walk. Since there is
little arithmetic intensity in these operations, the code is,
therefore, limited by the bandwidth of the device. We show in
\fig{\ref{fig:performance}} that our code achieves respectable
bandwidth: in excess of 50GB/s during the first tree walk, in which
only (cached) scatter memory reads are executed. The second tree walk,
which constructs the interaction list, is notably slower because there
data is written to memory--an operation which is significantly
slower compared to reads from texture memory.
We observe that the bandwidth decreases with
$\theta$ in both tree walks, which is due to increasingly divergent
tree-paths between neighbouring groups, and an increase of 
the write to read ratio in memory operations.

\section{Discussion and Conclusions}\label{sect:discussion}

We present a fast gravitational tree-code which is executed on Graphics Processing Units
(GPU) supporting the NVIDIA CUDA architecture. The novelty of this code is the GPU-based
tree-walk which, combined with the GPU-based calculation of accelerations, shows good
scaling for various particle numbers and different opening angles $\theta$. The hereby
produced energy error is comparable to existing CPU based tree-codes with quadrupole
corrections. The code makes optimal use of the available device resources and shows
excellent scaling to new architectures. Tests indicate that the NVIDIA GT200 architecture,
which has roughly twice the resources as the used G80 architecture, performs the
integration twice as fast. Specifically, the sustained science rate on a realistic galaxy
merger simulation with $8\cdot 10^5$ particles is $1.6\cdot 10^6$ particles/second. Our
tests revealed our GPU implementation to be two order of magnitudes faster than the
widely-used CPU version of the Barnes-Hut tree-code from the NEMO stellar dynamics package
\citep{1995ASPC...77..398T}. However, our code is only an order of magnitude faster 
compared to a SSE-vectorised tree-code specially tuned for x86-64 processors and 
the Phantom-grape library
\footnote{Private communication with Keigo Nitadori, the author of
  the Phtanom-grape library {\tt http://grape.mtk.nao.ac.jp/~nitadori/phantom/}}. Hamada et al
\citep{1654123} presented a similarly tuned tree code, in which Phantom-grape is replaced
with a GPU library \citep{2007astro.ph..3100H}. In this way, it was possible to achieve
the 200 GFLOP/s, and science rate of about $10^6$ particles/s. However, this code does not
include quadrupole corrections, and therefore GFLOP/s comparison between the two codes is
meaningless. Nevertheless, the science rate of the two codes is comparable for similar
opening angles, which implies that our tree-code provides more accurate results for the
same performance.

As it generally occurs with other algorithms, the introduction of a massively parallel
accelerator usually makes the host calculations and non-parallelisable parts of the code,
as small as they may be, the bottleneck. In our case, we used optimised device code and
for the host code we used general tree-construction and tree-walk recursive algorithms. It
is possible to improve these algorithms to increase the performance of the host part, but
it is likely to remain a bottleneck. Even with the use of modern quad-core processors this
part is hard to optimize since its largely a sequential operation.

\section*{Acknowledgements}
{\small
We thank Derek Groen, Stefan Harfst and Keigo
Nitadori for valuable suggestions and reading of the manuscript.  
This work is supported by NWO
(via grants \#635.000.303, \#643.200.503, VIDI grant \#639.042.607, 
VICI grant \#6939.073.803 and grant \#643.000.802). 
We thank the University of Amsterdam,
where part of this work was done, for their hospitality.
}


%
%

\section*{References}
\bibliographystyle{elsarticle-num}
\bibliography{GNMPZ08}







\end{document}